\documentclass[twocolumn,prd,showpacs,nofootinbib]{revtex4-1}
\usepackage[dvips]{epsfig}
\def\beq{\begin{equation}}
\def\eeq{\end{equation}}
\def\bea{\begin{eqnarray}}
\def\eea{\end{eqnarray}}
\def\bq{\begin{quote}}
\def\eq{\end{quote}}

\def\nnb{\nonumber}

\def\nnb{\nonumber}
\def\la{\langle}
\def\ra{\rangle}
\def\nin{\noindent}
\def\ba{\vspace*{-0.2cm}\begin{array}}
\def\ea{\end{array}\vspace*{-0.2cm}}

\def\als{\alpha_s}

\def\gg2{ \la\alpha_s G2 \ra}
\def\gg3{g^3f_{abc}\la G^aG^bG^c \ra}
\def\ggg4{\la\als^2G4\ra}

\def\beq{\begin{equation}}
\def\enq{\end{equation}}
\def\beqa{\begin{eqnarray}}
\def\enqa{\end{eqnarray}}
\def\nnb{\nonumber}

\def\MeV{\nobreak\,\mbox{MeV}}
\def\GeV{\nobreak\,\mbox{GeV}}

\def\qq{\lag\bar{q}q\rag}

\def\mix{\lag\bar{q}g\si.Gq\rag}

\def\G3{\lag g^3G3\rag}

\def\pli{p^\prime}

\def\si{\sigma}

\def\al{\alpha}

\def\lb{\label}
\def\nn{\nonumber}

\newcommand{\rag}{\rangle}
\newcommand{\lag}{\langle}

\begin{document}

\title{$Z_c^+(3900)$ decay width in QCD sum rules}

\author{J.M. Dias}
\email{jdias@if.usp.br}
 \author{F. S.  Navarra}
\email{navarra@if.usp.br}
%
\author{M. Nielsen}
\email{mnielsen@if.usp.br}
\affiliation{Instituto de F\'{\i}sica, Universidade de S\~{a}o Paulo,
C.P. 66318, 05389-970 S\~{a}o Paulo, SP, Brazil}
\author{C.M. Zanetti}
\email{carina.zanetti@gmail.com}
\affiliation{ Faculdade de Tecnologia, Universidade do Estado do Rio de 
Janeiro, Rod. Presidente Dutra Km 298, P\'olo Industrial, 27537-000 , 
Resende, RJ, Brasil}

\begin{abstract}
\nin
We identify the recently observed  charmonium-like structure $Z_c^\pm(3900)$
as the charged partner of the $X(3872)$ state. Using standard techniques of 
QCD sum rules, we evaluate the three-point function and extract the coupling 
constants of the $Z_c^+ \, J/\psi \, \pi^+$, $Z_c^+ \, \eta_c \, \rho^+$ and
$Z_c^+ \, D^+\bar{D}^{*0}$
vertices and the corresponding decay widths in these channels. The good 
agreement with the experimental data gives support to 
the tetraquark picture of this state.
\end{abstract}
\maketitle

\section{ Introduction}

About ten years after the discovery of the $X(3872)$, the BESIII collaboration 
has just reported the observation of a
charged charmonium-like structure in the $M(\pi^\pm J/\psi)$ mass spectrum 
of the $Y(4260)\to J/\psi\pi^+\pi^-$ decay channel \cite{Ablikim:2013mio}.
This structure, called $Z_c(3900)$, was also observed at the same time by 
BELLE \cite{Liu:2013dau} and was confirmed by the  authors of 
ref.~\cite{Xiao:2013iha} using  CLEO-c data.  
During the past decade, as other new  non-conventional states were discovered 
 their internal structure was subject of intense debate. Definite conclusions 
have not yet been reached and some models for these states are still under 
consideration: meson molecule \cite{swanson} , tetraquark \cite{maiani1}, 
hadro-charmonium \cite{voloshin}  and charmonium-molecule mixture 
\cite{zanetti}. For a comprehensive review of the theoretical and experimental 
status of these states we refer the reader to \cite{review}. 
In most of these models it is relatively easy to reproduce the masses of the 
states.  It is however much more difficult to reproduce their measured decay 
widths. In the present case, the  $Z_c(3900)$ decay width poses an additional  
challenge to theorists. Its mass is very close to  the $X(3872)$, which may be 
considered its neutral partner. However, while the $Z_c(3900)$ decay width is in 
the range $40-60$ MeV, the $X(3872)$ width is  smaller than 
$ 2.3 $ MeV.  A possible reason for this difference is the fact that the $X(3872)$ 
may contain a significant $|c \bar{c} \rangle$ component \cite{zanetti}, which is 
absent in the $Z_c(3900)$. Probably for this same reason the $Z_c$ was not observed 
in $B$ decays, as pointed out in \cite{zhao}.

In this work we present a calculation of the 
$Z_c(3900)$  decay width into $J/\psi \, \pi^+$,  $\eta_c \, \rho^+$ and
$Z_c^+ \, D^+\bar{D}^{*0}$.  

If the $Z_c$ is a real $D^* - \bar{D}$  molecular state its decay into  
$J/\psi \,  \pi^+$ (or $\eta_c \, \rho^+$) must involve the exchange of a charmed 
meson.  Since the 
exchange  of heavy mesons is a short range process, when the distance between  
$D^*$ and the  $ \bar{D}$ is large it becomes more difficult to exchange mesons.
 Using the expression of the decay width obtained with the one boson exchange 
potential  (OBEP), we can relate the decay width with the effective radius of 
the state. In \cite{namit} it was shown that, in order to reproduce the 
measured width, the effective radius must be $\langle r_{eff} \rangle \simeq 
0.4$ fm. This size scale is small and pushes the molecular picture to its 
limit of validity. In another work  \cite{hammer} the new state 
was again treated as a charged $D^* - \bar{D}$ molecule, in which the 
interaction between the charm mesons is described by a pionless effective 
field theory. Introducing electromagnetic interactions through the minimal 
substitution in this theory, the authors of \cite{hammer} were able to study 
the electromagnetic structure of the $Z_c$ and, in particular, its charge form 
factor and charge radius, which turned out to be $\langle r^2 \rangle \simeq 
0.11$ fm$^2$. Taking this radius as a measure of the spatial size of
the state, we conclude that it is more compact than a $J/\psi$, for which 
$\langle r^2 \rangle \simeq 0.16$ fm$^2$.  We take the combined results of 
\cite{namit} and \cite{hammer} as an indication that the $Z_c$ is a compact 
object, which may be better understood as a quark cluster, such as a 
tetraquark. Therefore in this work we explore this possibility.

As the number of new states increases, a new question arises concerning their 
grouping in families: which ones belong together? Which ones are  
groundstates and which are excitations?  
A possible organization of the charmonium and bottomonium  new states  was 
suggested in \cite{Navarra:2011xa} and it is summarized in Fig. 1. In the 
figure we compare the charm and bottom spectra in the mass region of interest. 
On the left (right) we show the charm (bottom) states with 
their mass differences in MeV. The comparison between the two left lines with 
the two lines on the right emphasizes the similarity between the spectra.  
In the bottom of the second column we have now 
the  newly found $Z_c(3900)$. In \cite{Navarra:2011xa} there was a question 
mark in this position.  In fact, the existence of a charged partner of the 
$X(3872)$ was first proposed in \cite{maiani1}. A few years later 
\cite{maiani} the same group proposed that the $Z^+(4430)$, observed by BELLE  
\cite{Belle:Z4430},  would be the first radial excitation of the charged
 partner of the $X(3872)$. This suggestion was based on the fact that the mass 
difference corresponding to a radial excitation in  the charmonium sector is 
given by $M_{\Psi (2S)} - M_{\Psi (1S)} = 590$ MeV. This number is  close to 
the mass difference $ M_{Z^+(4430)} - M_{X^+(3872)} = 560$ MeV. The very same 
connection between  $ Z^+(4430)$ and $Z_c(3900)$ was found in the hadro-charmonium 
approach \cite{voloshin2}, where the former is essentially a $\Psi^{\prime}$ embedded 
in light mesonic matter and the latter a $J/\psi$ also embedded in light mesonic matter. 
In a straightforward extension of this reasoning to the bottom sector, 
in \cite{Navarra:2011xa} it was conjectured that the $Z_b^+(10610)$, observed 
by the BELLE collaboration in \cite{bellezb},  may be a radial excitation of 
an yet unmeasured $X_b^+$. The observation of $Z_c^+(3900)$ gives support to 
this conjecture and should motivate  new experimental searches of this 
bottom charged state and its neutral partner, the only missing states in the 
diagram.

\begin{figure}[h]
\begin{center}
\centerline{\epsfig{figure=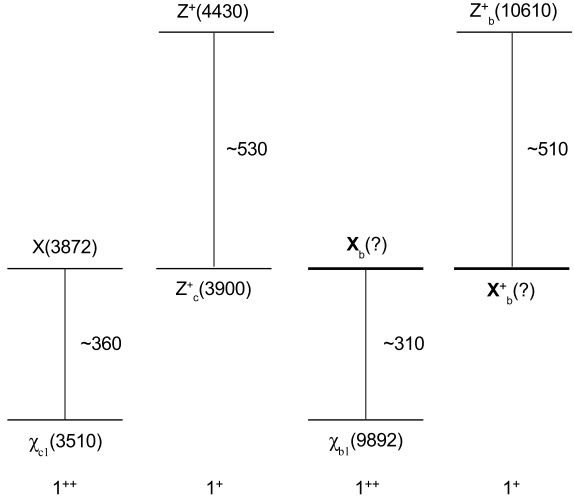,height=75mm}}
\caption{Charm and bottom energy levels in the mass region of interest. Masses 
are in MeV. On the two left columns we show the conjecture presented in 
\cite{maiani}. The $Z_c^+(3900)$ is  conjectured to be the charged
partner of the $X(3872)$. On the two right columns we show the conjecture 
advanced in \cite{Navarra:2011xa} for the bottom sector, where the $X_b(?)$ and 
$X_b^+(?)$ are the proposed states.}
\label{fig1}
\end{center}
\end{figure}

There are also other suppositions according to which the $Z_c^+(3900)$ should
be the charmed partner of the $Z_b^+(10610)$. In this scheme, there should exist
another charged state, called $Z_c^{'}$, that would be the charmed partner of the
$Z_b^+(10650)$ \cite{voloshin2,Faccini:2013lda,Guo:2013sya}.

In this work we use the  method of QCD  sum rules (QCDSR) \cite{svz,rry,SNB}
to study some hadronic decays of $Z_c(3900)$,
considering $Z_c$ as a four-quark state. 

\section{ $Z_c^+(3900) \to  J/\psi \, \pi^+$ Decay Width}

The QCDSR were used in ref.~\cite{x3872}
to study the  $X(3872)$ meson considered as a $I^G(J^{PC})=0^+(1^{++})$ 
four-quark state, and a good agreement with the experimental mass was obtained. 
The $Z_c(3900)$ is interpreted here as the isospin 1 partner of the $X(3872)$.
As in \cite{maiani,Faccini:2013lda} we assume the quantum numbers for the 
neutral state in the isospin multiplet to be $I^G(J^{PC})=1^+(1^{+-})$. 
Therefore, the interpolating field for $Z_c^+(3900)$ is given by:
\beq
j_\alpha={i\epsilon_{abc}\epsilon_{dec}\over\sqrt{2}}[(u_a^TC\gamma_5c_b)
(\bar{d}_d\gamma_\alpha C\bar{c}_e^T)-(u_a^TC\gamma_\alpha c_b)
(\bar{d}_d\gamma_5C\bar{c}_e^T)]\;,
\label{field}
\enq
where $a,~b,~c,~...$ are color indices, and $C$ is the charge conjugation
matrix. Considering $SU(2)$  symmetry, the mass obtained in QCDSR for the
$Z_c$ state is exactly the same one obtained  for the $X(3872)$, as it happens 
in the case of $\rho$ and $\omega$ states. There are also QCDSR
calculations for the $Z_c$ state considered as a $\bar{D}D^*$ molecular state
\cite{Cui:2013yva,Zhang:2013aoa}. These calculations only confirm the 
results presented in refs.~\cite{x3872,Narison:2010pd}.
Therefore here we evaluate only the decay width.

We start with the $Z_c^+(3900) \to  J/\psi \, \pi^+$ decay. 
The QCDSR calculation of the vertex  $Z_c(3900) \, J/\psi \,  \pi$ is based on 
the three-point function given by:
\beq
\Pi_{\mu\nu\al}(p,\pli,q)=\int d^4x~ d^4y ~e^{i\pli.x}~e^{iq.y}~
\Pi_{\mu\nu\al}(x,y),
\lb{3po}
\enq
with $\Pi_{\mu\nu\al}(x,y)=\lag 0 |T[j_\mu^{\psi}(x)j_{5\nu}^{\pi}(y)
j_\alpha^\dagger(0)]|0\rag$,
where $p=\pli+q$ and the interpolating fields for $J/\psi$ and $\pi$ 
are given by:
\beq
j_{\mu}^{\psi}=\bar{c}_a\gamma_\mu c_a,
\lb{psi}
\enq
\beq
j_{5\nu}^{\pi}=\bar{d}_a\gamma_5\gamma_\nu u_a,
\lb{pion}
\enq
In order to evaluate the phenomenological side of the sum rule we  
insert intermediate states for $Z_c$, $J/\psi$ and $\pi$ into Eq.(\ref{3po}). 
We get:
\beqa
\Pi_{\mu\nu\al}^{(phen)} (p,\pli,q)={\lambda_{Z_c} m_{\psi}f_{\psi}F_{\pi}~
g_{Z_c\psi \pi}(q^2)q_\nu
\over(p^2-m_{Z_c}^2)({\pli}^2-m_{\psi}^2)(q^2-m_\pi^2)}
\nn\\
~\left(-g_{\mu\lambda}+{\pli_\mu \pli_\lambda\over m_{\psi}^2}\right)
\left(-g_\alpha^\lambda+{p_\alpha p^\lambda\over m_{Z_c}^2}\right)
+\cdots\;,
\lb{phen}
\enqa
where the dots stand for the contribution of all possible excited states. 
The form factor, $g_{Z_c\psi \pi}(q^2)$, is defined as the generalization 
of the on-mass-shell matrix element, $\lag J/\psi \,  \pi \,| \, Z_c\rag$,
for an off-shell pion: 
\beq
\lag J/\psi(\pli) \pi(q)|Z_c(p)\rag=g_{Z_c\psi \pi}(q^2)
\varepsilon^*_\lambda(\pli)\varepsilon^\lambda(p),
\label{coup}
\enq
where
$\varepsilon_\alpha(p),~\varepsilon_\mu(\pli)$ are  the polarization
vectors of the $Z_c$ and $J/\psi$ mesons  respectively.
In deriving  Eq.~(\ref{phen}) we have used the definitions: 
\beqa
\lag 0 | j_\mu^\psi|J/\psi(\pli)\rag &=&m_\psi f_{\psi}\varepsilon_\mu(\pli),
\nn\\
\lag 0 | j_{5\nu}^\pi|\pi(q)\rag &=&iq_\nu F_\pi,
\nn\\
\lag Z_c(p) | j_\alpha|0\rag &=&\lambda_{Z_c}\varepsilon_\al^*(p).
\lb{fp}
\enqa

To extract directly the coupling constant, $g_{Z_c\psi \pi}$, instead
of the form factor, we can write a sum rule at the pion-pole \cite{Bracco:2011pg}, 
valid only at $Q^2=0$, as suggested in \cite{rry} for the pion-nucleon coupling
constant. This method was also applied to the nucleon-hyperon-kaon
coupling constant \cite{ccl,bnn} and to the nucleon$-\Lambda_c-D$ coupling
constant \cite{nn}. It consists in neglecting the pion mass in the denominator
of Eq.~(\ref{phen}) and working at $q^2=0$. In the OPE side only terms 
proportional to $1/q^2$ will contribute to the sum rule. Therefore, up to 
dimension five the only diagrams that contribute are the quark condensate
and the mixed condensate. 

\begin{figure}[h] \label{fig2}
\centerline{\epsfig{figure=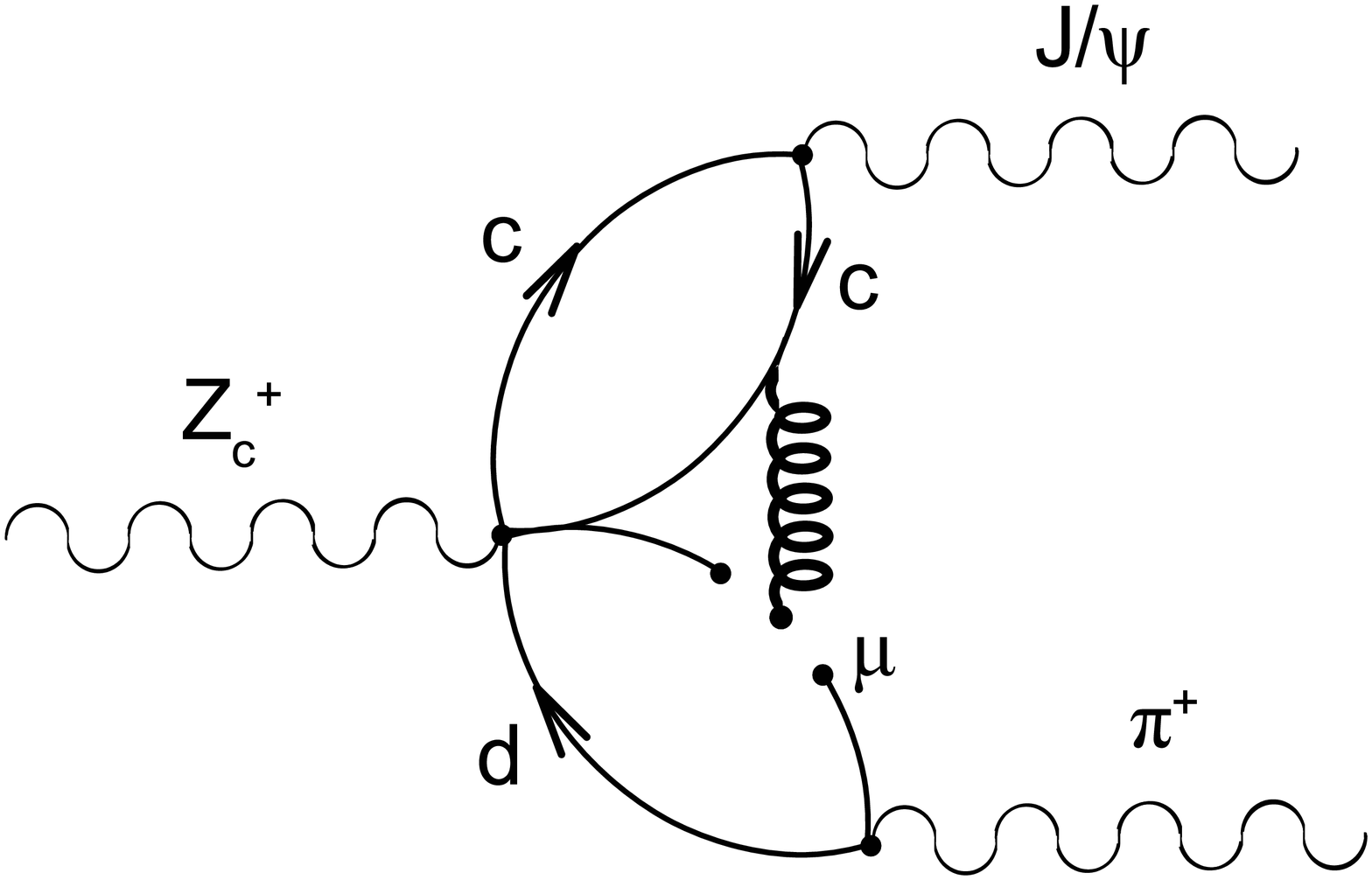,height=40mm}}
\caption{CC diagram which contributes to the OPE side of the sum rule.}
\end{figure} 

As discussed in refs.~\cite{Navarra:2006nd,ennr}, large partial decay widths are 
expected when the coupling constant is obtained from QCDSR in the case of 
multiquark states. By multiquark states we mean that the initial state contains 
the same number of valence quarks as the number of valence quarks in 
the final state. This happens  because, although the initial current,  
Eq.~(\ref{field}), has a non-trivial color structure, it can be rewritten as a sum 
of molecular type currents with trivial color configuration through a
 Fierz transformation. To avoid this problem we follow
refs.~\cite{Navarra:2006nd,ennr}, and consider in the OPE side only the diagrams
with non-trivial color structure, which are called color-connected (CC) diagrams.
In the present case the CC diagram that contributes to the OPE side at
the pion pole is shown  in Fig.~2.  Possible permutations (not shown) of the 
diagram in Fig.~2 also contribute.

The diagram in Fig.~2 contributes only to the structures $q_\nu g_{\mu\alpha}$
and $q_\nu\pli_\mu\pli_\alpha$ appearing in the phenomenological side. Since 
structures with more momenta are supposed to give better results, we choose 
to work with
the $q_\nu\pli_\mu\pli_\alpha$ structure. Therefore  in the OPE side and in
the $q_\nu\pli_\mu\pli_\alpha$ structure we obtain:
\beq
\Pi^{(OPE)}= {\mix\over12\sqrt{2}\pi^2}{1\over q^2}
\int_0^1 d\alpha{\alpha(1-\al)\over m_c^2-\al(1-\al){\pli}^2}.
\label{ope}
\enq
Isolating the $q_\nu\pli_\mu\pli_\alpha$ structure in Eq.~(\ref{phen}) and 
making a single Borel transformation to both $P^2={P^\prime}^2\rightarrow M^2$,
we finally get the sum rule:
\beqa
A\left(e^{-m_\psi^2/M^2}-e^{-m_{Z_c}^2/M^2}\right)+B~e^{-s_0/M^2}=
\nn\\
={\mix\over12\sqrt{2}\pi^2}
\int_0^1 d\alpha \,  e^{- m_c^2\over \al(1-\al)M^2},
\label{sr}
\enqa
where $s_0$ is the continuum threshold parameter for $Z_c$,
\beq
A={g_{Z_c\psi \pi}\lambda_{Z_c} f_{\psi}F_{\pi}~(m_{Z_c}^2+m_\psi^2)
\over 2m_{Z_c}^2m_{\psi}(m_{Z_c}^2-m_{\psi}^2)},
\label{a}
\enq
and $B$ is a parameter introduced to take into account single pole 
contributions associated with pole-continuum transitions,
which are not suppressed  when only a single Borel transformation
is done in a three-point function sum rule \cite{Navarra:2006nd,col,bel,io1}.
In the numerical analysis we use the following values for quark masses and QCD 
condensates \cite{x3872,narpdg}:
\beqa\label{qcdparam}
&m_c(m_c)=(1.23\pm 0.05)\,\GeV,\nnb\\
&\qq = - (0.23\pm0.03)^3
\,\GeV^3,\nnb\\
&\lag\bar{q}g\si.Gq\rag=m_0^2\lag\bar{q}q\rag,\nnb\\
&m_0^2=
0.8\,\GeV^2.
\enqa
For the meson masses and decay constants we use the experimental values 
\cite{pdg} $m_{\psi}=3.1$ GeV, $m_\pi=138$ MeV, $f_{\psi}=0.405$ GeV and 
$F_\pi=131.52$ MeV. For the $Z_c$ mass we use the value measured in  
\cite{Ablikim:2013mio}: $m_{Z_c}=(3899\pm6)$ MeV. The meson-current coupling, 
$\lambda_{Z_c}$, defined in Eq.(\ref{fp}), can be determined from the two-point 
sum rule \cite{x3872}: $\lambda_{Z_c}=(1.5\pm0.3)\times10^{-2}~\GeV^5$. 
For the continuum threshold we use $s_0=(m_{Z_c}+\Delta s_0)^2$, with $\Delta s_0=
(0.5\pm0.1)~\GeV$.
\begin{figure}[h] \label{fig3}
\centerline{\epsfig{figure=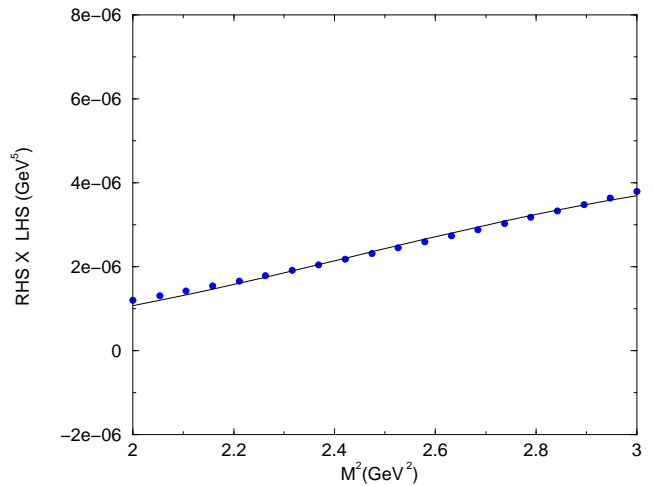,height=65mm}}
\caption{Dots: the RHS of Eq.(\ref{sr}), as a function of the Borel mass
for $\Delta s_0=0.5$ GeV.  
The solid line gives the fit of the QCDSR results through 
the LHS of Eq.(\ref{sr}).}
\end{figure} 
We evaluate the sum rule in the range $2.0 \leq M^2 \leq 3.0$ GeV$^2$,  
which is the range where the two-point function for $X(3872)$ (which is the same 
for $Z_c(3900)$) shows good OPE convergence and where the pole contribution 
is bigger than the continuum contribution \cite{x3872}. 
In Fig.~3 we show, through the circles, the right-hand side (RHS) of 
Eq.(\ref{sr}), as a function of the Borel mass.

To determine the coupling constant $g_{Z_c\psi \pi}$ we fit the QCDSR results 
with the analytical expression in the left-hand side (LHS) of Eq.(\ref{sr}),
and find (using $\Delta s_0=0.5~\GeV$): $A=1.46\times10^{-4}~\GeV^5$ and 
$B=-8.44\times10^{-4}~\GeV^5$. 
Using the definition of $A$ in Eq.(\ref{a}), the value obtained for the
coupling constant is $g_{Z_c\psi \pi}=3.89~\GeV$, which is in excellent agreement 
with the estimate made in \cite{Faccini:2013lda}, based on  dimensional arguments.
Considering the uncertainties given above, we finally find:
\beq
g_{Z_c\psi \pi}=(3.89\pm0.56)~\GeV.
\label{coupling}
\enq

The decay width is given by \cite{Faccini:2013lda}:
\beqa
\Gamma(Z_c^+(3900)\to J/\psi\pi^+)={p^*(m_{Z_c},m_\psi,m_\pi)\over8\pi m_{Z_c}^2}
\nn\\
\times{1\over3}g^2_{Z_c\psi \pi}\left(
3+{(p^*(m_{Z_c},m_\psi,m_\pi))^2\over m_{\psi}^2}\right),
\label{dec}
\enqa
where 
\beq
p^*(a,b,c)={\sqrt{a^4+b^4+c^4-2a^2b^2-2a^2c^2-2b^2c^2}\over 2a}.
\enq 
\vskip0.5cm
Therefore we obtain:
\beq
\Gamma(Z_c^+(3900)\to J/\psi\pi^+)=(29.1\pm8.2)~\MeV.
\label{width}
\enq
\section{ $Z_c^+(3900) \to  \eta_c \, \rho^+$ Decay Width}

Next we consider the $Z_c^+(3900)\to \eta_c \, \rho^+$ decay. The three-point
function for the  corresponding vertex is obtained from
Eq.~(\ref{3po}) by using
\beq
\Pi_{\mu\al}(x,y)=\lag 0 |T[j_5^{\eta_c}(x)j_{\mu}^{\rho}(y)
j_\alpha^\dagger(0)]|0\rag,
\enq 
with
\beq
j_{5}^{\eta_c}=i\bar{c}_a\gamma_5 c_a,\mbox{ and }
j_{\mu}^{\rho}=\bar{d}_a\gamma_\mu u_a.
\lb{rho}
\enq
In this case the phenomenological side is
\beqa
&&\Pi_{\mu\al}^{(phen)} (p,\pli,q)={-i\lambda_{Z_c} m_{\rho}f_{\rho}f_{\eta_c}
m^2_{\eta_c}~g_{Z_c\eta_c \rho}(q^2)
\over2m_c(p^2-m_{Z_c}^2)({\pli}^2-m_{\eta_c}^2)(q^2-m_\rho^2)}
\nn\\
&\times&\left(-g_{\mu\lambda}+{q_\mu q_\lambda\over m_{\rho}^2}\right)
\left(-g_\alpha^\lambda+{p_\alpha p^\lambda\over m_{Z_c}^2}\right)
+\cdots,
\lb{phen2}
\enqa
where now we have used the definitions: 
\beqa
\lag 0 | j_\mu^\rho|\rho(q)\rag &=&m_\rho f_{\rho}\varepsilon_\mu(q),
\nn\\
\lag 0 | j_{5}^{\eta_c}|\eta_c(\pli)\rag &=& {f_{\eta_c}m^2_{\eta_c}\over2m_c}.
\lb{fp2}
\enqa
In the OPE side we consider the CC diagrams of the same kind of the diagram in 
Fig.~2. In the $\pli_\alpha q_\mu$ structure we have:
\beq
\Pi^{(OPE)}= {-i m_c\mix\over48\sqrt{2}\pi^2}{1\over q^2}
\int_0^1 d\alpha{1\over m_c^2-\al(1-\al){\pli}^2}.
\label{ope2}
\enq
Remembering that $p = \pli + q $, isolating the $q_\al\pli_\mu$ structure in 
Eq.~(\ref{phen2}) and 
making a single Borel transformation on  both $P^2={P^\prime}^2\rightarrow M^2$,
we finally get the sum rule:
\beqa
C\left(e^{-m_{\eta_c}^2/M^2}-e^{-m_{Z_c}^2/M^2}\right)+D~e^{-s_0/M^2}=
\nn\\
{Q^2+m_\rho^2\over Q^2}{m_c\mix\over48\sqrt{2}\pi^2}
\int_0^1 d\alpha {e^{- m_c^2\over \al(1-\al)M^2}\over\al(1-\al)},
\label{sr2}
\enqa
with $Q^2=-q^2$ and
\beq
C={g_{Z_c\eta_c \rho}(Q^2)\lambda_{Z_c} m_\rho f_{\rho}f_{\eta_c}m_{\eta_c}^2
\over2m_c m_{Z_c}^2(m_{Z_c}^2-m_{\eta_c}^2)}.
\label{c}
\enq
We use the experimental values for $ m_\rho,~ f_{\rho}$ and $m_{\eta_c}$ 
\cite{pdg} and we extract $f_{\eta_c}$ from ref.~\cite{nov}:
\beqa
m_\rho=0.775~\GeV,\;\;m_{\eta_c}=2.98~\GeV,
\nn\\
f_{\rho}=0.157~\GeV,\;\;f_{\eta_c}=0.35~\GeV.
\enqa
\begin{figure}[h] 
\centerline{\epsfig{figure=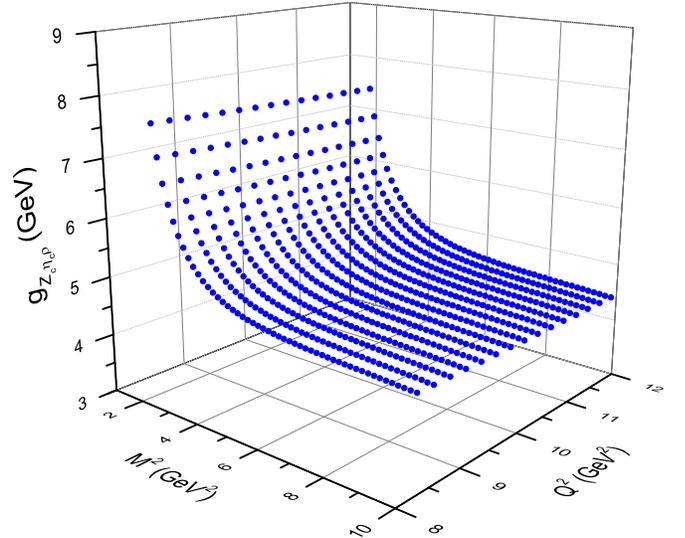,height=80mm}}
\caption{QCDSR results for the form factor  $g{Z_c\eta_c \rho}(Q^2)$ as 
a function of $Q^2$ and $M^2$ for $\Delta s_0=0.5$ GeV.}
\label{fig4}
\end{figure} 
One can use Eq.~(\ref{sr2}) and its derivative with respect to $M^2$ to 
eliminate $D$ from Eq.~(\ref{sr2}) and to isolate $ g_{Z_c\eta_c \rho}(Q^2)$. 
In Fig. \ref{fig4} we show $g_{Z_c\eta_c \rho}(Q^2)$ as a function of
both $M^2$ and $Q^2$. A good Borel window is determined when the parameter to be 
extracted from the sum rule is as much independent of the Borel mass
as possible. Therefore, from Fig. \ref{fig4} we notice that 
the Borel window where the form factor is independent of $M^2$ is in the region 
$4.0 \leq M^2 \leq 10.0$ GeV$^2$.
\begin{figure}[h]
\centerline{\epsfig{figure=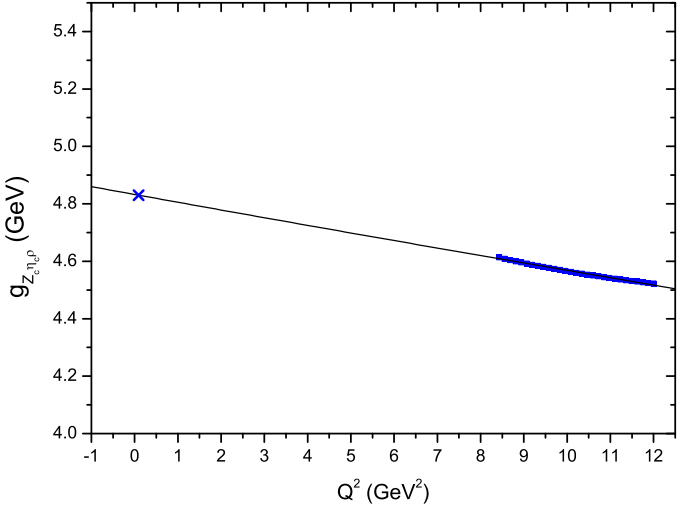,height=80mm}}
\caption{QCDSR results for $g_{Z_c\eta_c \rho}(Q^2)$, as a function of $Q^2$, 
for $\Delta s_0=0.5$ GeV (squares). 
The solid line gives the parametrization of the QCDSR results  through Eq. 
(\ref{exp}). The cross gives the value of the coupling constant.}
\label{fig5} 
\end{figure} 
The squares in Fig. \ref{fig5} show the $Q^2$ dependence 
of $g_{Z_c\eta_c \rho}(Q^2)$, obtained for $M^2=5.0$ GeV$^2$. For 
other values of the Borel mass, in  the range  
$4.0 \leq M^2 \leq 10.0$ GeV$^2$, the results are equivalent.
Since the coupling constant is defined as the value of the form factor at the 
meson pole: $Q^2 = -m^2_{\rho}$,  we need to extrapolate 
the form factor for a region of $Q^2$ where
the QCDSR are not valid. This extrapolation 
can be done by parametrizing the QCDSR results for 
$g_{Z_c\eta_c\rho}(Q^2)$ with the help of an exponential form:

\beq
g_{Z_c\eta_c\rho}(Q^2) = g_1e^{-g_2 Q^2},
\label{exp}
\enq
with $g_1 = 4.83$ GeV and $ g_2=5.6\times10 ^{-3}~\mbox{GeV}^{-2}$.  
We also show in Fig. \ref{fig5}, through the line, the fit of the QCDSR results
for $\Delta s_0=0.5$ GeV, using Eq.~(\ref{exp}). The value of the 
 coupling constant, $g_{Z_c\eta_c\rho}$, is also shown in this figure through
the cross. We obtain:

\beq
g_{Z_c\eta_c\rho}=g_{Z_c\eta_c\rho}(-m^2_\rho)=(4.85 \pm 0.81)~~\mbox{GeV}.
\label{couprho}
\enq
The uncertainty in the coupling constant  given above comes from 
variations in $s_0$, $\lambda_{Z_c}$ and $m_c$ in the ranges given above.
This value for the coupling is bigger than the estimate presented in 
\cite{Faccini:2013lda}. Inserting this coupling and the corresponding masses
into Eq.~(\ref{dec}) we find

\beq
\Gamma(Z_c^+(3900)\to \eta_c\rho^+)=(27.5\pm8.5)~\MeV.
\label{width2}
\enq
\section{ $Z_c^+(3900) \to  D^+ \bar{D}^{*0}$ Decay Width}

Finally we consider the $Z_c^+(3900)\to D^+\bar{D}^{*0}$ decay. In this case we 
use in Eq.~(\ref{3po})
\beq
\Pi_{\mu\al}(x,y)=\lag 0 |T[j_\mu^{D^*}(x)j_5^{D}(y)
j_\alpha^\dagger(0)]|0\rag,
\enq 
where
\beq
j_{5}^{D}=i\bar{d}_a\gamma_5 c_a,\mbox{ and }
j_{\mu}^{D^*}=\bar{c}_a\gamma_\mu u_a.
\lb{D}
\enq
Using the definitions
\beqa
\lag 0 | j_\mu^{D^*}|D^*(\pli)\rag &=&m_{D^*} f_{D^*}\varepsilon_\mu(\pli),
\nn\\
\lag 0 | j_{5}^{D}|D(q)\rag &=& {f_{D}m^2_{D}\over m_c},
\lb{fdd*}
\enqa
the phenomenological side is given by
\beqa
&&\Pi_{\mu\al}^{(phen)} (p,\pli,q)={-i\lambda_{Z_c} m_{D^*}f_{D^*}f_{D}
m^2_{D}~g_{Z_cDD^*}(q^2)
\over m_c(p^2-m_{Z_c}^2)({\pli}^2-m_{D^*}^2)(q^2-m_D^2)}
\nn\\
&\times&\left(-g_{\mu\lambda}+{\pli_\mu \pli_\lambda\over m_{D^*}^2}\right)
\left(-g_\alpha^\lambda+{p_\alpha p^\lambda\over m_{Z_c}^2}\right)
+\cdots.
\lb{phen3}
\enqa

In the OPE side we consider again only the CC diagrams. In the 
$\pli_\alpha \pli_\mu$ structure we have:
\beqa
\Pi^{(OPE)}&=& {-i m_c\mix\over48\sqrt{2}\pi^2}\left[{1\over m_c^2-q^2}
\int_0^1 d\alpha{\alpha(2+\alpha)\over m_c^2-(1-\al){\pli}^2}\right.
\nn\\
&-&\left.{1\over m_c^2-{\pli}^2}
\int_0^1 d\alpha{\alpha(2+\alpha)\over m_c^2-(1-\al)q^2}\right].
\label{ope3}
\enqa
\begin{figure}[h] 
\centerline{\epsfig{figure=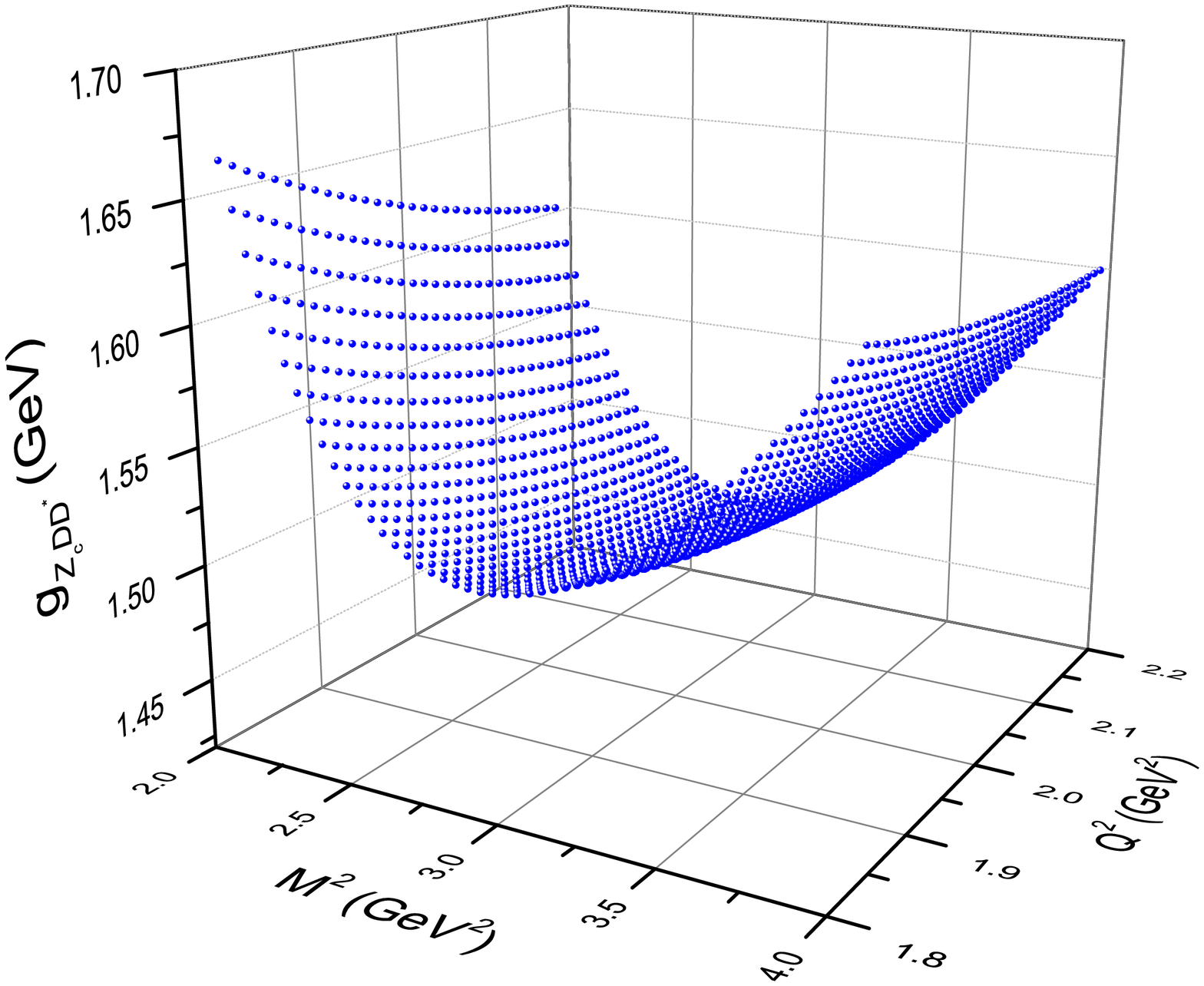,height=80mm}}
\caption{QCDSR results for the form factor  $g_{Z_cDD^*}(Q^2)$ as 
a function of $Q^2$ and $M^2$ for $\Delta s_0=0.5$ GeV.}
\label{fig6}
\end{figure} 
Isolating the $\pli_\mu\pli_\alpha$ structure in 
Eq.~(\ref{phen3}) and 
making a single Borel transformation on  both $P^2={P^\prime}^2\rightarrow M^2$,
we get:
\beqa
&&{1\over Q^2+m_D^2}\left[E\left(e^{-m_{D^*}^2/M^2}-e^{-m_{Z_c}^2/M^2}\right)+
F~e^{-s_0/M^2}\right]=
\nn\\
&&{m_c\mix\over48\sqrt{2}\pi^2}\left[{1\over m_c^2+Q^2}
\int_0^1 d\alpha{\alpha(2+\alpha)\over1-\al}~ e^{- m_c^2\over \al(1-\al)M^2}\right.
\nn\\
&-&\left.e^{-m_c^2/M^2}
\int_0^1 d\alpha{\alpha(2+\alpha)\over m_c^2+(1-\al)Q^2}\right],
\label{sr3}
\enqa
with 
\beq
E={g_{Z_cDD^*}(Q^2)\lambda_{Z_c} f_{D^*}f_{D}m_{D}^2
\over m_cm_{D^*}(m_{Z_c}^2-m_{D^*}^2)}.
\label{e}
\enq

We use the experimental values for $ m_D$ and $m_{D^*}$ 
\cite{pdg} and we extract $f_{D}$ and $f_{D^*}$ from ref.~\cite{Bracco:2011pg}:
\beqa
m_D=1.869~\GeV,\;\;f_{D}=(0.18\pm0.02)~\GeV,
\nn\\
m_{D^*}=2.01~\GeV,\;\;f_{D^*}=(0.24\pm0.02)~\GeV.
\enqa
In Fig. \ref{fig6} we show $g_{Z_cDD^*}(Q^2)$, as a function of
both $M^2$ and $Q^2$, from where we notice that we get a Borel stability
in the region $2.2 \leq M^2 \leq 2.8$ GeV$^2$.

\begin{figure}[h]
\centerline{\epsfig{figure=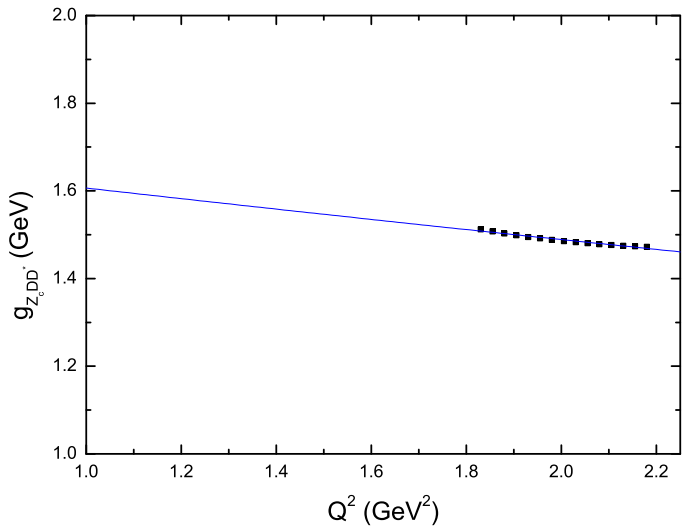,height=70mm}}
\caption{QCDSR results for $g_{Z_cDD^*}(Q^2)$, as a function of $Q^2$, 
for $\Delta s_0=0.5$ GeV (squares). 
The solid line gives the parametrization of the QCDSR results  through Eq. 
(\ref{exp}).}
\label{fig7} 
\end{figure} 


Fixing $M^2=2.6$ GeV$^2$ we show in Fig. \ref{fig7}, through the squares, 
the $Q^2$ dependence of the $g_{Z_cDD^*}(Q^2)$ form factor. Again, to extract 
the coupling constant we fit the QCDSR results
 using the exponential form in Eq.~(\ref{exp}) with $g_1 = 1.733$ GeV and 
$ g_2=0.076~\mbox{GeV}^{-2}$.  
The line in in Fig. \ref{fig7} shows the fit of the QCDSR results
for $\Delta s_0=0.5$ GeV, using Eq.~(\ref{exp}). 
We get for the coupling constant:

\beq
g_{Z_cDD^*}=g_{Z_cDD^*}(-m^2_D)=(2.5 \pm 0.3)~~\mbox{GeV}.
\label{coupdd}
\enq
The uncertainty in the coupling constant  comes from 
variations in $s_0$, $\lambda_{Z_c},~f_D,~f_{D^*}$ and $m_c$.
This value for this coupling is again in excelent agreement with  the estimate 
presented in \cite{Faccini:2013lda}. Using again Eq.~(\ref{dec}) with this coupling, 
the decay width in this channel is

\beq
\Gamma(Z_c^+\to D^+\bar{D}^{*0})=(3.2\pm0.7)~\MeV.
\label{width3}
\enq

\section{Conclusions}

In conclusion, we have used the three-point QCDSR to evaluate the coupling
constants in the vertices $Z_c^+(3900)J/\psi\pi^+$, $Z_c^+(3900)\eta_c\rho^+$ and
$Z_c^+(3900)D^+\bar{D}^{*0}$.
In the case of the $Z_c^+(3900)J/\psi\pi^+$ vertex, we have used the sum rule at 
the pion pole, and the coupling was extracted directly from the sum rule. 
In the cases of $Z_c^+(3900)\eta_c\rho^+$ and $Z_c^+(3900)D^+\bar{D}^{*0}$ 
vertices, we have extracted the form 
factors, and the couplings were obtained with a fit of the QCDSR results. In the
three  cases  we have only considered the color connected diagrams, since we 
expect  the
$Z_c(3900)$ to be a genuine tetraquark state with a non-trivial color 
structure.
The obtained couplings, with the respective decay widths, are given in Table I.
We have also included in this table the results for the vertex 
$Z_c^+(3900)\bar{D}^0{D}^{*+}$, since it is exactly the same result as in the
$Z_c^+(3900)D^+\bar{D}^{*0}$ vertex. 

\begin{center}
\small{{\bf Table I:} Coupling constants and decay widths in  different 
channels.}
\\
\vskip3mm
\begin{tabular}{c|c|c}  \hline
 Vertex& coupling constant (GeV) & decay width (MeV)\\
\hline
 $Z_c^+(3900)J/\psi\pi^+$ & $3.89\pm0.56$ & $29.1\pm8.2$ \\
 $Z_c^+(3900)\eta_c\rho^+$ & $4.85 \pm 0.81$ & $27.5\pm8.5$ \\
 $Z_c^+(3900)D^+\bar{D}^{*0}$ & $2.5 \pm 0.3$ & $3.2\pm0.7$ \\
 $Z_c^+(3900)\bar{D}^0D^{*+}$ & $2.5 \pm 0.3$ & $3.2\pm0.7$ \\
\hline
\end{tabular}\end{center}

 Considering these 
four decay channels we get a total width $\Gamma=(63.0\pm 18.1)$ GeV for 
$Z_c(3900)$ which is in agreement with the two experimental values:
$\Gamma=(46\pm 22)$ MeV from BESIII \cite{Ablikim:2013mio}, and
$\Gamma=(63\pm35)$ MeV from BELLE \cite{Liu:2013dau}.

 \subsection*{Acknowledgments}
 \noindent
This work has been supported by  CNPq and FAPESP-Brazil.


\begin{thebibliography}{99}

\bibitem{Ablikim:2013mio} 
  M.~Ablikim {\it et al.}  [BESIII Collaboration],  arXiv:1303.5949.

\bibitem{Liu:2013dau} 
  Z.Q.~Liu {\it et al.}  [BELLE Collaboration],  arXiv:1304.0121.

\bibitem{Xiao:2013iha} 
  T.~Xiao, S.~Dobbs, A.~Tomaradze and K.K.~Seth,  arXiv:1304.3036.


\bibitem{swanson} F.E. Close and P.R. Page,  Phys. Lett. B {\bf 628}, 215 (2005); 
E. Braaten and M. Kusunoki, Phys. Rev.  D {\bf 69}, 074005 (2004); F.E. Close and 
P.R. Page, Phys. Lett. B {\bf 578}, 119 (2004); 
	          N.A. Tornqvist, Phys. Lett. B {\bf 590}, 209 (2004); 
                  E.S. Swanson, Phys. Rept. {\bf 429}, 243 (2006); 
                  S. Fleming, M. Kusunoki, T. Mehen, and U. van Kolck, Phys. Rev. 
D {\bf 76}, 034006 (2007); E. Braaten and M. Lu, Phys. Rev. D {\bf 76}, 094028 
(2007). 


\bibitem{maiani1}  L.~Maiani, F.~Piccinini, A.D.~Polosa and V.~Riquer,   
Phys.\ Rev.\ D {\bf 71}, 014028 (2005).


\bibitem{voloshin}   S.~Dubynskiy and M.B.~Voloshin,   Phys.\ Lett.\ B 
{\bf 666}, 344 (2008).



\bibitem{zanetti}   R.D'E. Matheus, F.S. Navarra, M. Nielsen and C.M. Zanetti,
 Phys.\ Rev.\ D {\bf 80}, 056002 (2009).

\bibitem{review}   N.~Brambilla {\it et al.}, Eur.\ Phys.\ J.\ C {\bf 71}, 
1534 (2011); M.~Nielsen, F.S.~Navarra and S.H.~Lee,  Phys.\ Rept.\  {\bf 497},
 41 (2010), and  references therein. 

\bibitem{zhao}  Q.~Wang, C.~Hanhart and Q.~Zhao,  arXiv:1303.6355 [hep-ph].


\bibitem{namit} N. Mahajan, arXiv:1304.1301 [hep-ph].

\bibitem{hammer}   E.~Wilbring, H.-W.~Hammer and U.-G.~Meißner,
                   arXiv:1304.2882.

\bibitem{Navarra:2011xa}
  F.S.~Navarra, M.~Nielsen and J.M.~Richard,
  J.\ Phys.\ Conf.\ Ser.\  {\bf 348}, 012007 (2012) [arXiv:1108.1230].


\bibitem{maiani}  L.~Maiani, A.D.~Polosa and V.~Riquer,
  arXiv:0708.3997.

\bibitem{Belle:Z4430}
  S.K.~Choi {\it et al.}  [Belle Collaboration],
  Phys.\ Rev.\ Lett.\  {\bf 100}, 142001 (2008).



\bibitem{voloshin2}   M.B.~Voloshin,  arXiv:1304.0380.


\bibitem{bellezb} I. Adachi {\it et al.} [BELLE Collaboration],
arXiv:1105.4583.
 
\bibitem{Faccini:2013lda} 
  R. Faccini, L. Maiani, F. Piccinini, A. Pilloni, A.D. Polosa and V. Riquer,
   arXiv:1303.6857.

\bibitem{Guo:2013sya} 
  F.-K.~Guo, C.~Hidalgo-Duque, J.~Nieves and M.P.~Valderrama,  arXiv:1303.6608.

\bibitem{svz} M.A. Shifman, A.I. and Vainshtein and V.I. Zakharov,
Nucl. Phys. B {\bf 147}, 385 (1979).

\bibitem{rry} L.J. Reinders, H. Rubinstein and S. Yazaki, Phys. Rept.
{\bf 127}, 1 (1985).

\bibitem{SNB} For a review and references to original works, see
e.g., S.
Narison, {\it QCD as a theory of hadrons,
Cambridge Monogr. Part. Phys. Nucl. Phys. Cosmol.} {\bf 17}, 1 (2002)
[hep-h/0205006]; {\it QCD
spectral sum rules ,  World Sci. Lect. Notes Phys.} {\bf 26}, 1 (1989);
{ Acta Phys. Pol.} B  {\bf 26}, 687 (1995); { Riv. Nuov. Cim.} {\bf 10N2}, 1
(1987); { Phys. Rept.} {\bf 84}, 263 (1982).

\bibitem{x3872} 
  R.D'E.~Matheus, S.~Narison, M.~Nielsen and J.M.~Richard,
  Phys.\ Rev.\ D {\bf 75}, 014005 (2007) [hep-ph/0608297].


\bibitem{Cui:2013yva} 
  C.-Y.~Cui, Y.-L.~Liu, W.-B.~Chen and M.-Q.~Huang,  arXiv:1304.1850.

\bibitem{Zhang:2013aoa} 
  J.-R.~Zhang,  arXiv:1304.5748.

\bibitem{Narison:2010pd}   S.~Narison, F.S.~Navarra and M.~Nielsen,
  Phys.\ Rev.\ D {\bf 83}, 016004 (2011) [arXiv:1006.4802].
 
\bibitem{Bracco:2011pg} 
  M.E.~Bracco, M.~Chiapparini, F.S.~Navarra and M.~Nielsen,
  Prog.\ Part.\ Nucl.\ Phys.\  {\bf 67}, 1019 (2012) [arXiv:1104.2864].

\bibitem{ccl} S. Choe, M.K. Cheoun and S.H. Lee, { Phys. Rev.} C {\bf 53},
              1363 (1996); S. Choe, { Phys. Rev.} C {\bf 57},
              2061 (1998).

\bibitem{bnn} M.E. Bracco, F.S. Navarra and M. Nielsen, 
               { Phys. Lett.} B {\bf 454}, 346 (1999).

\bibitem{nn}   F.S. Navarra and M. Nielsen, 
               { Phys. Lett.} B {\bf 443}, 285 (1998).

\bibitem{Navarra:2006nd} 
  F.S.~Navarra and M.~Nielsen,  Phys.\ Lett.\ B {\bf 639}, 272 (2006); 
  F.O.~Duraes, S.H.~Lee, F.S.~Navarra and M.~Nielsen,  Phys.\ Lett.\ 
B {\bf 564}, 97 (2003). 

\bibitem{ennr} M. Eidem\"uller {\it et al.},  Phys.\ Rev.\ D {\bf 72}, 034003 
(2005)  [hep-ph/0503193].

\bibitem{col} P. Colangelo {\it et al.}, { Phys. Lett.} B {\bf 339}, 151 (1994).

\bibitem{bel} V.M. Belyaev {\it et al.}, { Phys. Rev.} D {\bf 51}, 6177 (1995). 

\bibitem{io1} B.L. Ioffe and A.V. Smilga, { Nucl. Phys.} B {\bf 232},  109
(1984).

\bibitem{narpdg} S. Narison, Phys. Lett. B {\bf 466}, 345 (1999);
S. Narison,  Phys. Lett. B {\bf 361}, 121 (1995);
S. Narison, Phys. Lett. B {\bf 387}, 162 (1996); S. Narison,  Phys.  Lett. B
{\bf 624}, 223 (2005).

\bibitem{pdg} J. Beringer {\it et al.} (Particle Data Group), Phys. Rev. D 
{\bf 86}, 010001 (2012).

\bibitem{nov} V.A. Novikov {\it et al.}, Phys. Rep. {\bf 41}, 1 (1978); N.G.
Deshpande and J. Trampetic,  Phys. Lett. B {\bf 339}, 270 (1994).

\end{thebibliography}
\end{document}